# Discrete Charging in Polysilicon Gates of Single Electron Transistors

Dharmraj Kotekar-Patil, Stefan Jauerneck, David Wharam, Dieter Kern
Institute for Applied Physics, University of Tübingen, Germany
Xavier Jehl, Romain Wacquez, M. Sanquer
SPSMS, UMR-E CEA/UJF-Grenoble 1, INAC, and CEA/LETI-MINATEC  Grenoble F-38054, France
e-mail: david.wharam@uni-tuebingen.de

*Abstract*—Low temperature electron transport measurements of single electron transistors fabricated in advanced CMOS technology with polysilicon gates not only exhibit clear Coulomb blockade behavior but also show a large number of additional conductance fluctuations in the nonlinear regime. By comparison with simulations these features are quantitatively attributed to the effects of discretely charged islands in the polysilicon gates.

## I. Introduction

Single electron transistors (SETs) are very sensitive to charge fluctuations in their vicinity and hence are excellent charge detectors [1]. Different techniques are used to define a nanoscale island with tunnel barriers in Silicon based materials, e.g. doped single crystal silicon wires with constrictions [2,3,4], polysilicon nanowires [5,6,7], and MOSFET structures [8,9,10]. In polysilicon wire devices, grain boundaries form tunnel barriers due to trap states and individual grains may act as islands, whereas in MOSFET structures islands and barriers are formed by doping modulation along the length of the wire. In this paper we present low-temperature measurements on small MOS-SETs where the charging events in the polysilicon gate are detected by the SET. The associated conductance fluctuations have previously been interpreted as arising from interplay between Coulomb interaction, valley splitting, and strong quantum confinement [11]. Here we present an alternative explanation for the origin of these conductance features as well as a simple electrostatic model which explains all the features observed in the experimental data.

## II. Fabrication

Our samples are fabricated on 200 mm silicon-on-insulator (SOI) wafers utilizing the CMOS technology platform at CEA-LETI [12]. A nanowire is etched from 8 nm thick silicon, a 5 nm gate oxide is grown and a gate is then etched from a deposited polysilicon layer of typically 50nm thickness (Fig. 1). Silicon nitride spacers of 25 nm width are formed on both sides of the gate. Heavy As implantation ($2 \times 10^{15}$ cm$^{-2}$) is performed to form self-aligned source and drain regions using the gate and spacers as a mask. In our devices, the region below the gate forms a quantum dot upon application of a positive voltage to the gate, while the region below the spacers forms barriers between the dot and the adjacent source and drain reservoirs.

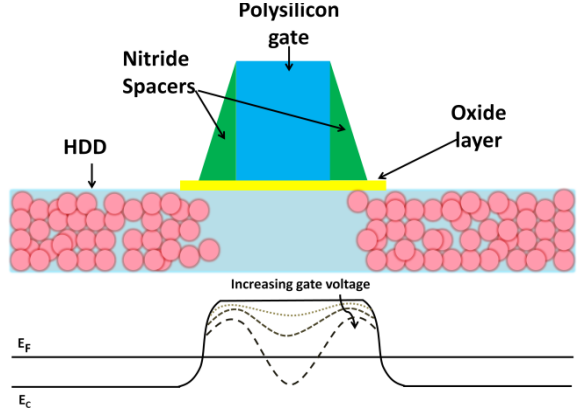

Figure 1. Top: Schematic of the device. Bottom: The conduction band profile for increasing gate voltage is shown and demonstrates the formation of an island separated from source and drain by tunnel barriers.

### III. MEASUREMENTS

All the measurements reported here were performed in a dilution refrigerator operating at a base temperature of 70 mK. Two methods were used for the measurements. In one (differential conductance), a small ac excitation voltage, typically 50 µV, was applied between source and drain reservoirs and the conductance through the device was measured using low-frequency (27.3 Hz) phase-sensitive current detection. In the second method (transconductance) a small ac excitation was applied to the gate electrode and conductance through the device was measured, again using low-frequency phase-sensitive detection.

Fig. 2a shows the result of differential conductance measurements where clear Coulomb blockade diamonds are visible due to the dot formed below the gate (D1 in fig. 3). However, in the conducting regions the conductance is modulated in a regular pattern of parallel lines of high and low differential conductance. Such lines are also clearly visible in the transconductance measurements in fig. 2b.

These conductance lines have a positive slope, $dV_d/dV_g \approx 2$ and a typical period of $\Delta V_g = 1mV$. The slopes of these lines are unusual because they contradict the basic model of Coulomb charging of a single island. The non-linear conductance for an island tunnel coupled to source and drain, and capacitively coupled to a gate electrode is expected to show the generic Coulomb diamonds with positive edge slopes $|dV_d/dV_g|$ always less than unity. The signature of processes such as excited state transport is expected to run parallel to the edges. A possible explanation for the structure in the conductance data is the existence of an additional island (D2 in fig. 3). The small period of these additional lines suggests that the size of D2 is much larger than the expected size of D1. In the differential conductance measurements, the conductance lines exhibit positive differential conductance (PDC) along the edge of the Coulomb diamond with positive slope, and negative differential conductance (NDC) along the negative slope edge of the Coulomb diamond. In the transconductance measurements we observe the opposite behaviour. The lines exhibit negative transconductance (NTC) along the Coulomb diamond edge with positive slope and positive transconductance (PTC) along the edge with negative slope. We also observe that at the junction of successive Coulomb diamonds, where we expect the electrochemical potentials of source, D1 and drain to align, (marked by circle in fig. 2a) the blockade is not lifted. Furthermore we expect D2 to be purely capacitively coupled to source and drain, otherwise we would expect a finte conductance within the blockade of $D_1$ which is not observed in our measurements. This suggests that the origin of the observed lines is extrinsic to the SET. It is also interesting to note that such lines are absent in devices with silicided gates [13]. We attribute these lines to the charging of polysilicon grains within the gate (D2 in fig. 3). At sufficiently low temperatures, if the charging energy of the polysilicon grain is larger than $k_B T$ and the resistance of the grain boundaries, $R_g$ larger than $h/e^2$, the quantum of resistance, the charging of the polysilicon grains is dominated by the Coulomb blockade [14].

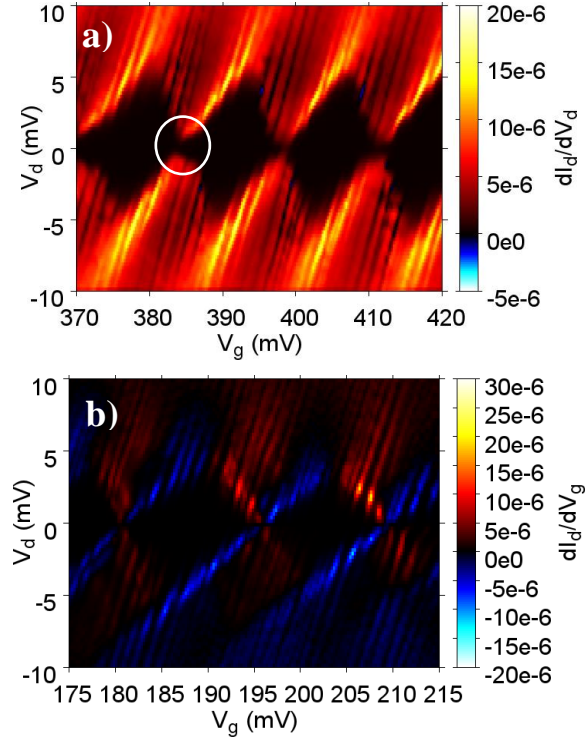

Figure 2. a) Differential conductance measurement at a bath temperature of 70 mK. Measurement is done by applying a small ac signal of 50 µV to source-drain (frequency = 27.3 Hz). b) Transconductance measurement at 70 mK done with small ac signal applied to the gate (50 µV and frequency = 27.3 Hz). In both cases lines of higher and lower conductance and transconductance, respectively, with slopes of 2 are visible.

## IV. ELECTROSTATIC MODEL

We therefore propose an electrostatic model based on the conclusions drawn from these measurements. It resembles a single electron box (SEB) capacitively coupled to an SET [15] where D1 is tunnel coupled to source and drain and capacitively coupled to D2, whereas D2 is capacitively coupled to source, drain and D1 and tunnel coupled to the gate (fig. 3). Assuming that D2 is much larger than D1, the effect of charging D1 seen by D2 is comparatively small. So for simplicity we may solve the electrostatic problem for the SEB independently from the SET. The electrochemical potential of D2 is given by [14] (for the notation please refer to fig. 3):

$$\mu_{D2} = \frac{(eN_2)^2}{C_{\Sigma 2}} - \frac{e(C_{S2}V_s + C_{D2}V_d + C_{G2}V_g)}{C_{\Sigma 2}}$$

where $C_{\Sigma 2} = C_{S2} + C_{D2} + C_{G2}$, assuming that $C_{dd}$ is very small compared to $C_{S2}$, $C_{D2}$, $C_{G2}$.

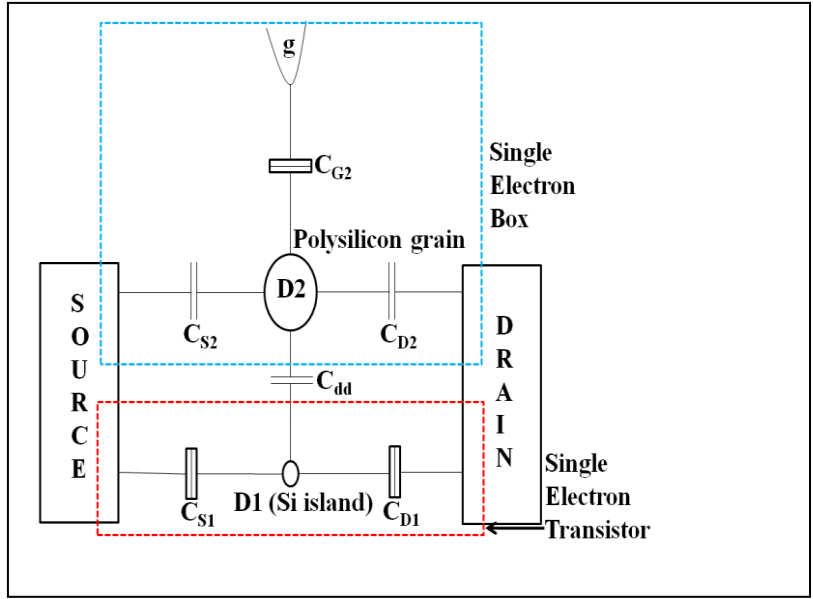

Figure 3. Schematic of a polysilicon grain (D2) capacitively coupled to an SET.

Since D2 is tunnel-coupled to the gate, the number of electrons $N_2$ on D2 changes by 1 when the electrochemical potential of the gate $\mu_g = \mu_{g0}-eV_g$ (where $\mu_{g0}$ is the electrochemical potential of the gate with no gate voltage applied) aligns with the electrochemical potential of D2 ($\mu_{D2}$).

Therefore, upon application of a gate bias, the condition required to charge D2 with an electron is:

$$\mu_{g0}-eV_g = \frac{(eN_2)^2}{C_{\Sigma 2}} - \frac{e(C_{S2}V_s + C_{D2}V_d + C_{G2}V_g)}{C_{\Sigma 2}}$$

Rearranging this equation and differentiating with respect to $V_g$ gives us the slope of the boundary line in the $V_g$-$V_d$ plane where the charge on D2 fluctuates.

$$\frac{dV_d}{dV_g} = \frac{C_{D2}+C_{S2}}{C_{D2}}$$

Although we assume $C_{dd}$ to be negligible when considering the charging of the grain through the gate, it still couples the SET to the SEB as is seen in the conductance measurements. For a symmetric capacitive coupling of D2 to source and drain a slope of 2 is to be expected in the conductance measurements (Fig. 2a and 2b), as is indeed observed in our measured data. Hence the conductance measurements of the SET reveal the charging of the SEB.

Using the capacitative model shown in fig. 3 transport simulations have been performed based upon the master equation technique [16, 17]. The parameters used for the simulation are summarized in table. 1. Fig. 4a shows the simulated current $I_d$ through the SET as a function of the source-drain voltage $V_d$ and the voltage $V_g$ applied to the gate g. The $I_d(V_g,V_d)$ surface shows distinct steps along lines with slope 2 in the $V_g,V_d$ -plane. They are the result of the discrete discharging of the polysilicon grain $D_2$ with increasing gate voltage $V_g$.

| Capacitor | Capacitance (aF) | Barrier Conductance ($e^2/h$) |
|---|---|---|
| $C_{S1}$ | 11.4 | 0.1 |
| $C_{D1}$ | 9 | 0.1 |
| $C_{S2}$ | 80 | 0 |
| $C_{D2}$ | 80 | 0 |
| $C_{dd}$ | 14 | 0 |
| $C_{G2}$ | 100 | 0.1 |

Table 1. Parameters used for simulations in fig. 4a - fig. 4c.

This can be understood if we again for simplicity consider the charging behavior of the single electron box neglecting its effect on the SET. Then one can easily show that the voltage across $C_{dd}$, the effective gate voltage $V_{dd}$ of the SET, is given by

$$V_{dd} = \tilde{V}(V_s, V_d) + \frac{C_{G2}}{C_{\Sigma 2}} V_g - \frac{N_2 e}{C_{\Sigma 2}}$$

With the parameters from table 1 this results in a jump of $V_{dd}$ by ~ 0.6 mV each time $N_2$ decreases by 1, i.e. an electron leaves the polysilicon grain.

By considering the $I_d(V_g, V_d)$-trend along the blue line in fig. 4a one can conclude that differential conductance measurements should show lines of positive differential conductance along the edges of the Coulomb diamond with positive slope, and negative differential conductances lines along an edge with negative slope. Similarly we can conclude by following the yellow line that transconductance measurements should exhibit the opposite behavior: lines of negative transconductance along an edge with positive slope and positive transconductance along an edge with negative slope.

Fig. 4b shows the simulated differential conductance and fig. 4c shows the simulated transconductance for the model shown in fig. 3. Simulated data in fig. 4b and fig. 4c exhibit conductance lines with slope = 2 in good agreement with experimental data shown in fig. 2a and 2b, respectively. Only conductance lines with positive slopes are visible in the measured and simulated data resulting from charging of D2. Since D2 is connected to only one tunnel barrier, the one to the gate, the charge on D2 can fluctuate only when $\mu_g = \mu_{D2}$ which results in conductance lines with positive slopes.

The conclusions of our simple model have been confirmed by our more detailed simulations which take the capacitive coupling between the SET and the SEB into account. Both these simulations and our detailed differential and trans- conductance measurements show positive and negative conductance lines along the slopes of the Coulomb diamonds. If we consider the junction of successive Coulomb diamonds where the electrochemical potentials of the source, D1 and drain align, a discrete change of the charge state of D2 pushes the electrochemical potential of D1 outside the transport window, bringing the dot D1 back into the blockade region. As a result the blockade is not lifted for a finite amount of bias (fig. 2a).

a)

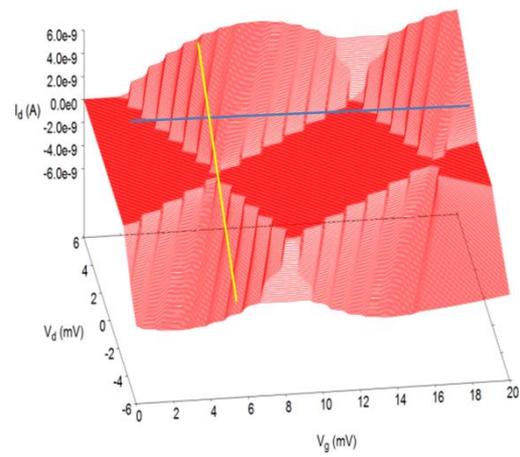

b)

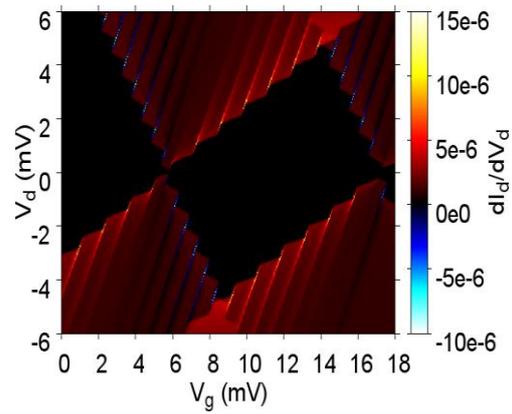

c)

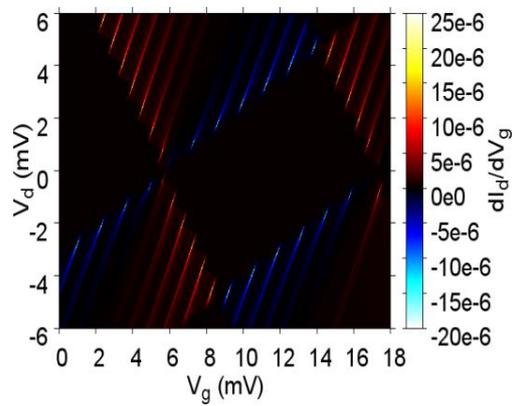

Figure 4. a) Simulated current through the SET based on the model in fig. 3 showing steps along lines with slope = 2 (symmetric case, $C_{S2}=C_{D2}$). b) Simulated differential conductance for this device. c) Simulated transonductance.

## V. Conclusion

We have shown that in small MOSFETS with polysilicon gates at very low temperatures the charging of an individual grain in the polysilicon can influence the device conductance, as evidenced by operating the device as an SET. The slopes of the observed lines in the conductance measurements as well as their spacing and the appearance of positive and negative differential conductance can be explained by a simple electrostatic model.

## Acknowledgement

The research leading to these results has received funding from the European Community's seventh Framework (FP7 2007/2013) under the Grant Agreement Nr: 214989. The samples in this work have been designed and fabricated by the AFSID Project Partners (http://www.afsid.eu). Authors would like to thank Mathieu Pierre for providing the simulation program.